\documentclass[twocolumn]{aastex63}

\usepackage{xspace}
\usepackage{textcomp}

\newcommand{\fref}{Fig.~\ref}
\newcommand{\eref}{Eq.~\ref}

\newcommand\nustar{\textit{NuSTAR}\xspace}
\newcommand\nicer{\textit{NICER}\xspace}
\newcommand\swift{\textit{Swift}\xspace}
\newcommand\us{\ensuremath{\mathrm{\mu sec}}\xspace}

\graphicspath{{./}{figures/}}

\received{January 1, 2018}
\revised{January 7, 2018}
\accepted{\today}

\submitjournal{ApJ}

\shorttitle{\nustar timing calibration}
\shortauthors{Bachetti et al.}

\begin{document}

\title{Timing Calibration of the \nustar X-ray Telescope}

\author[0000-0002-4576-9337]{Matteo Bachetti}
\altaffiliation{Fulbright Visiting Scholar}
\affiliation{INAF-Osservatorio Astronomico di Cagliari, via della Scienza 5, I-09047 Selargius, Italy}
\affiliation{Space Radiation Laboratory, Caltech, 1200 E California Blvd, Pasadena, CA 91125}
\email{matteo.bachetti@inaf.it}

\author[0000-0001-9803-3879]{Craig B. Markwardt}
\affiliation{Astrophysics Science Division, NASA Goddard Space Flight Center, Greenbelt, MD 20771, USA}

\author[0000-0002-1984-2932]{Brian W. Grefenstette}
\affiliation{Space Radiation Laboratory, Caltech, 1200 E California Blvd, Pasadena, CA 91125}

\author[0000-0003-3847-3957]{Eric V. Gotthelf}
\affiliation{Columbia Astrophysics Laboratory, Columbia University, 550 West 120th Street, New York, NY 10027-6601, USA}

\author[0000-0002-7889-6586]{Lucien Kuiper}
\affiliation{SRON Netherlands Institute for Space Research, Sorbonnelaan 2, NL-3584 CA Utrecht, the Netherlands}

\author[0000-0002-0393-9190]{Didier Barret}
\affiliation{IRAP, Université de Toulouse, CNRS, UPS, CNES, 9, Avenue du Colonel Roche, BP 44346, F-31028 Toulouse Cedex 4, France}

\author{W. Rick Cook}
\affiliation{Space Radiation Laboratory, Caltech, 1200 E California Blvd, Pasadena, CA 91125}

\author[0000-0002-9922-8915]{Andrew Davis}
\affiliation{Space Radiation Laboratory, Caltech, 1200 E California Blvd, Pasadena, CA 91125}

\author[0000-0003-0388-0560]{{Felix F\"urst}}
\affiliation{European Space Astronomy Centre (ESA/ESAC), Operations Department, Villanueva de la Cañada (Madrid), Spain}

\author[0000-0001-5800-5531]{Karl Forster}
\affiliation{Space Radiation Laboratory, Caltech, 1200 E California Blvd, Pasadena, CA 91125}

\author[0000-0003-2992-8024]{Fiona A. Harrison}
\affiliation{Space Radiation Laboratory, Caltech, 1200 E California Blvd, Pasadena, CA 91125}

% \author[0000-0001-9345-0307]{Victoria M. Kaspi}
% \affiliation{Department of Physics, McGill University, Montreal, Quebec, Canada}
% \affiliation{McGill Space Institute, McGill University, Montreal, Quebec, Canada}

\author[0000-0003-1252-4891]{Kristin K. Madsen}
\affiliation{Space Radiation Laboratory, Caltech, 1200 E California Blvd, Pasadena, CA 91125}
\affiliation{Astrophysics Science Division, NASA Goddard Space Flight Center, Greenbelt, MD 20771, USA}

\author[0000-0002-8074-4186]{Hiromasa Miyasaka}
\affiliation{Space Radiation Laboratory, Caltech, 1200 E California Blvd, Pasadena, CA 91125}

\author{Bryce Roberts}
\affiliation{Space Sciences Laboratory, 7 Gauss Way, University of California, Berkeley, CA 94720-7450, USA}

\author[0000-0001-5506-9855]{John A. Tomsick}
\affiliation{Space Sciences Laboratory, 7 Gauss Way, University of California, Berkeley, CA 94720-7450, USA}

\author[0000-0001-5819-3552]{Dominic J. walton}
\affiliation{Institute of Astronomy, University of Cambridge, Madingley Road, Cambridge CB3 0HA, UK}

% \author{Many More people\texttrademark}

\begin{abstract}
The Nuclear Spectroscopic Telescope Array (\nustar) mission is the first focusing X-ray telescope in the hard X-ray (3-79 keV) band.
Among the phenomena that can be studied in this energy band, some require high time resolution and stability: rotation-powered and accreting millisecond pulsars, fast variability from black holes and neutron stars, X-ray bursts, and more.
Moreover, a good alignment of the timestamps of X-ray photons to UTC is key for multi-instrument studies of fast astrophysical processes.
In this Paper, we describe  the timing calibration of the \nustar mission. 
In particular, we present a method to correct the temperature-dependent frequency response of the on-board temperature-compensated crystal oscillator.

Together with measurements of the spacecraft clock offsets obtained during downlinks passes, this allows a  precise   characterization of the behavior of the oscillator. The calibrated \nustar event timestamps for a typical observation are shown to be accurate to a precision of $\sim65\,\us$.

\end{abstract}


\section{Introduction} \label{sec:intro}
The \textit{Nuclear Spectroscopic Telescope ARray}  \citep[\nustar][]{harrison_nuclear_2013} is the first focusing hard (3--79\,keV) X-ray mission.
Compared to other missions covering the same energy range, \nustar provided a 10-fold improvement in angular resolution (58$''$ HPD), while also granting good spectral resolution (0.4\,keV at 6\,keV), and high effective area ($\sim1000$\,cm$^2$ at 10\,keV).
\nustar was designed with an absolute timing accuracy requirement of 100 ms \citep{harrison_nuclear_2013}.
Typically, the uncertainties in the time-stamping of \nustar data are much smaller than this requirement \citep{madsen_calibration_2015}: most electronic and propagation delays can be tracked down to $\sim100$\,\us precision. 
The largest unmodeled issue in the \nustar timing calibration has been, until now, a $\sim$2 ms drift of the spacecraft clock which is only tracked accurately between ground passes, spaced by a few hours.
\citet{gotthelf2017} showed that the drift could be tracked by using fast millisecond pulsars with very sharp pulse profiles, and that on short time scales the time measurement was stable enough to show $\sim15 \ \us$-thin pulsar peaks.
However, the actual process creating the drift remained unknown, limiting the range of timing capabilities of \nustar.

In this Paper, we show that this drift is largely due to a temperature-dependent frequency drift of the spacecraft temperature-compensated quartz oscillator (TCXO).
Our detailed modeling of this behavior and the clock aging allows the timing calibration of \nustar to be improved by almost two orders of magnitude.
This opens up \nustar to a wide range of new applications involving rapid variability, like rotation- and accretion-powered millisecond pulsars, X-ray bursts, and kHz quasi-periodic oscillations \citep{lorimerBinaryMillisecondPulsars2008,vanderklisRapidXrayVariability2006}, or precise synchronization with other satellites for multi-instrument and/or multiwavelength studies.

We describe the \nustar time tagging procedure in Section~\ref{sec:timetag}. 
Then, we model the temperature dependence of the TCXO in Section~\ref{sec:fvst}.
We discuss the improved timing calibration in Section~\ref{sec:precision}.

\section{\nustar time tagging details}\label{sec:timetag}

%The time tagging of \nustar events is described in detail in \citet{madsen_calibration_2015,bachetti_no_2015}.
%The precise time tagging of X-ray events done by a satellite like \nustar requires us to model accurately all sources of uncertainty in the electronics, the stability of the reference clock in the spacecraft, the propagation delays during the downlink of the data, and the precision of the reference clock in the ground station.

The method used to assign photon arrival times on the spacecraft for \nustar detector events is described in detail in \citet{madsen_calibration_2015,bachetti_no_2015}.
Photon arrival timestamps are assigned to received photons using
the on-board spacecraft oscillator. The spacecraft oscillator 
itself is calibrated to UT using ground station ranging measurements.
Converting photon arrival times to UT requires an accurate model of the uncertainties in the on-board electronics, the stability of the spacecraft reference clock, the propagation delays during the downlink of the data, and the precision of the reference clock in the ground station.

The systematic electronic delays on board are modeled to a few \us; the propagation delay to the ground station is on the order of 300\, ms and is modeled accurately ($<$100 \us) thanks to frequent tracking of \nustar's geographic location\footnote{\nustar uses the same downlink procedure as \swift, that was validated to the $\sim 100\,\us$ level \citep{cusumanoTimingAccuracySwift2012}}; the Malindi ground station, nominally used to measure \nustar's clock offsets, is synchronized to UTC via GPS with a precision of a few nanoseconds. 
This is not true of the less frequently used USN Hawaii and Singapore ground stations, that guarantee only 1-ms precision (see below).

However, the largest source of uncertainty in the \nustar clock modeling is the stability of the spacecraft's oscillator.
%temperature-compensated quartz oscillator.
%(TCXO). 
The mission does not carry onboard a GPS-synchronized clock. 
Event time stamps are referred to the pulse-per second (PPS) signal coming from a MC623X2 TCXO.
The nominal frequency (or \textit{divisor}) of this oscillator is $\sim$24\,MHz, meaning that the oscillator ticks $\sim$24 million times before sending the next PPS, a number that can be adjusted remotely through a command to the spacecraft. 
% The nominal temperature stability of this oscillator is between $\pm$2 and $\pm4$ parts per million (ppm) over the frequency interval -40--85\,C.
However, this frequency changes by $\sim$1\,ppm with the change of the oscillator temperature during an orbit. 
This produces an accumulated delay between the recorded time and the reference UTC time from the ground station. 

Every time \nustar connects to a ground station for the downlink of the data, the ground station measures the relative departure of the clock time from UT. 
If the clock has drifted more than an acceptable limit (chosen to be about 100\,ms), the divisor of the spacecraft TCXO is adjusted in order to change the direction of the drift.
This is done to satisfy the mission requirement of 100\,ms timing precision.
%All clock offset measurements and divisor changes are recorded and are made available to the \nustar Science Operations Center (hereafter SOC). The \nustar SOC uses these clock offsets to produce a clock-correction file, distributed as part of the \nustar CALDB from HEASARC and usable with the \texttt{barycorr} FTOOL.
The clock offset measurements and divisor changes are used by the \nustar Science Operations Center (hereafter SOC) to produce the clock-correction files, which are distributed as part of the \nustar CALDB from HEASARC and usable with the FTOOL \texttt{barycorr} needed to convert spacecraft UT time to the Solar system barycenter.

%However, the temperature changes rapidly during each of the $\sim 97$ minutes of a \nustar orbit, and so does the TCXO frequency;  the delay between the recorded times from UT changes too rapidly to be accurately tracked using ground passes spaced by a few hours. 

However, the temperature changes rapidly during each of the $\sim 97$ minutes of a \nustar orbit, which is found to effect the TCXO frequency with a similar $\sim97$\,minute modulation. 
Because of the relatively infrequent ground contacts, the time offset measurements do not accurately reflect the instantaneous arrival times of the events as registered by the spacecraft.
As a consequence, the clock correction file produced at the \nustar SOC is only able to model the drift to the $\sim 2\,$ms level.

In the following sections, we characterize the frequency-temperature relation in detail, allowing a much better approximation of the clock drift.

\section{Temperature and aging characterization of the TCXO}\label{sec:fvst}
\begin{figure*}
\includegraphics[width=3.5in]{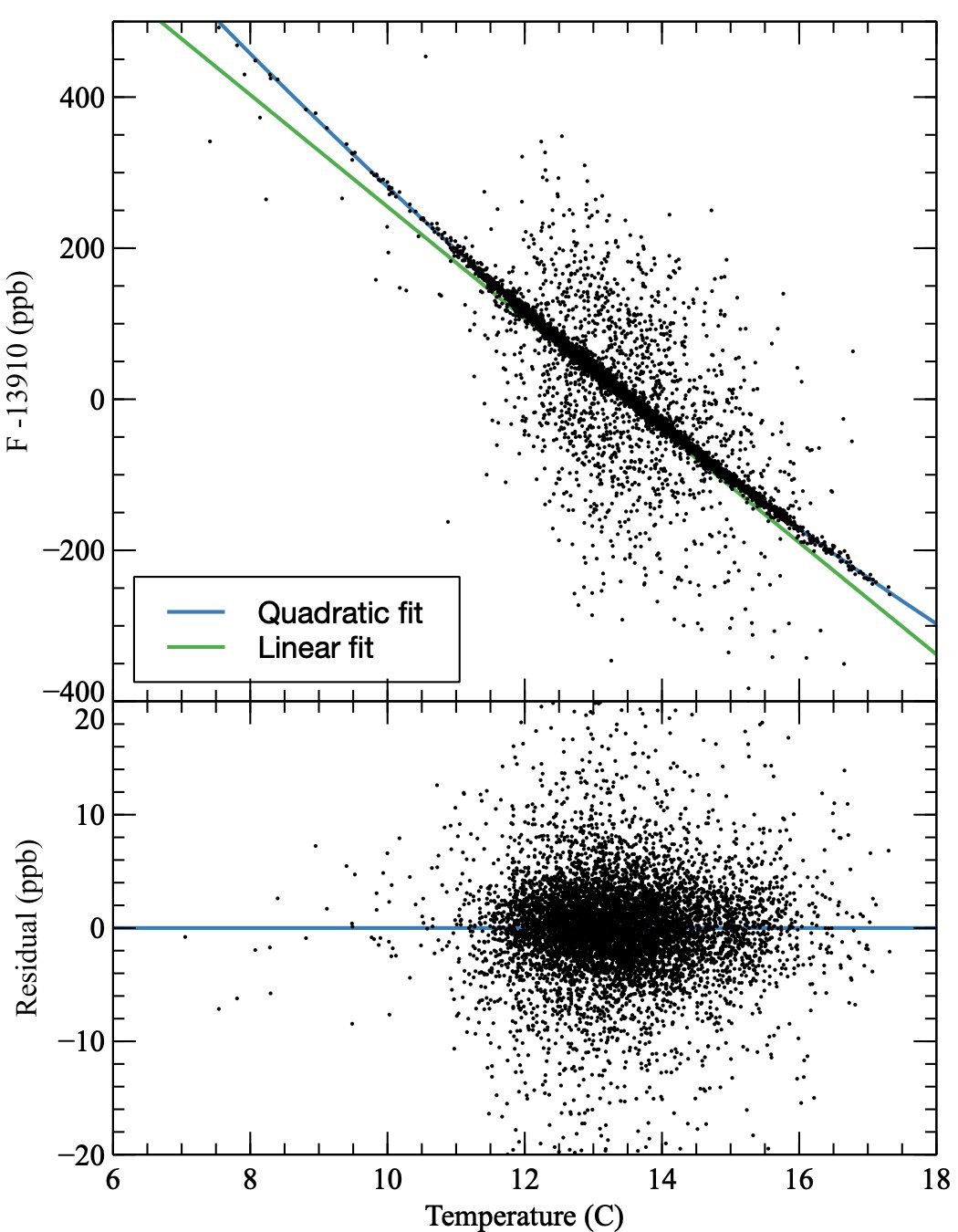}
\includegraphics[width=3.5in]{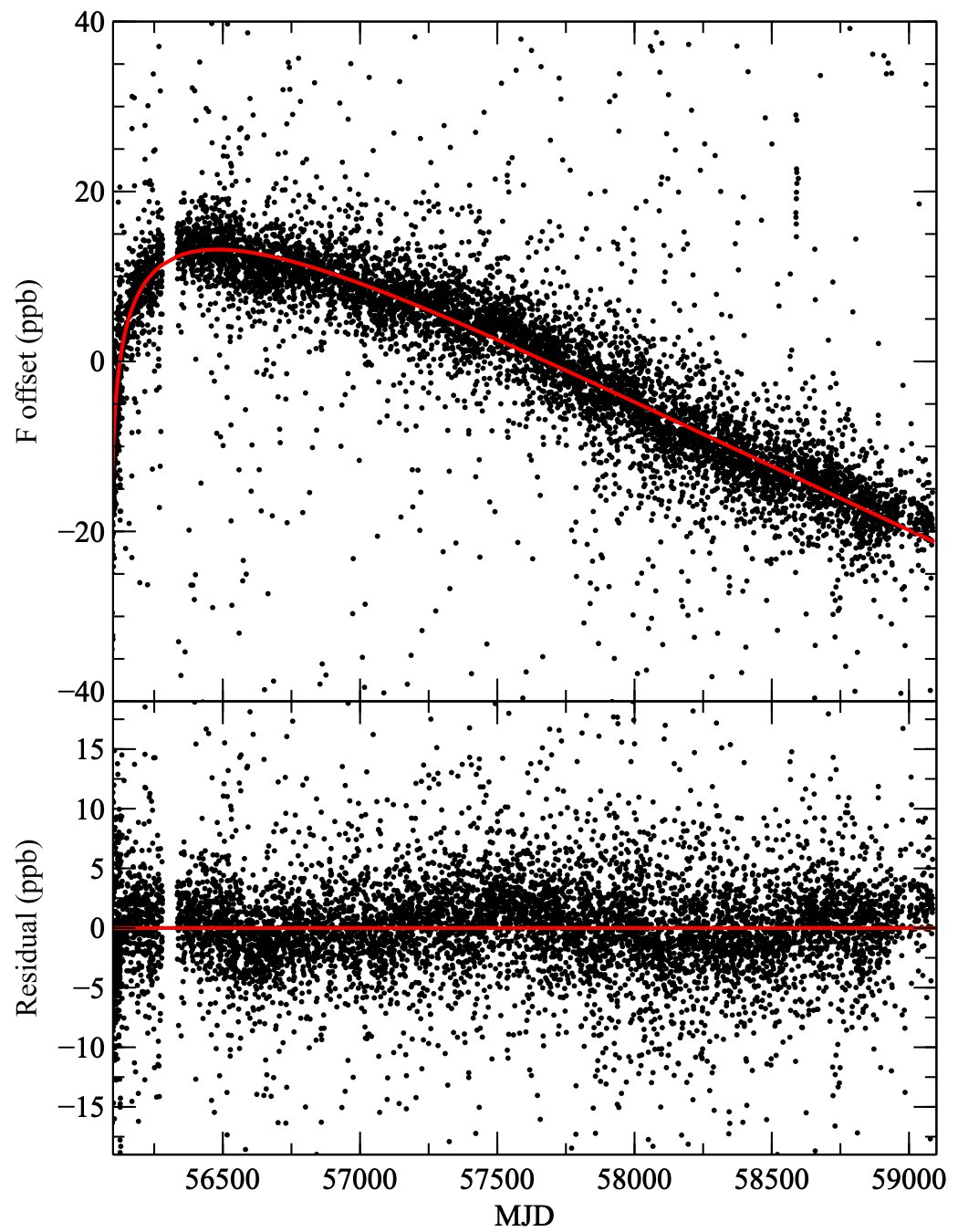}
\caption{Correction (in \textit{parts per billion}, pbb) for the long-term degradation of the spacecraft's crystal oscillator clock over time. 
(Left) A quadratic fit to the clock frequency in the f0-T plane after correcting for aging effects as described in the text. 
The blue line shows the quadratic fit to \eref{eq:modelA}, while the green line shows the linear component of the model to guide the eye;
(Right) The aging of the crystal oscillator corrected for temperature effects;
Bottom panels show the residuals from the above models, whose scatter is less than 10 parts per \textit{billion}.
\label{fig:f0ppm}}
\end{figure*}

\subsection{Basic modeling framework}
\begin{figure*}
\includegraphics[width=\linewidth]{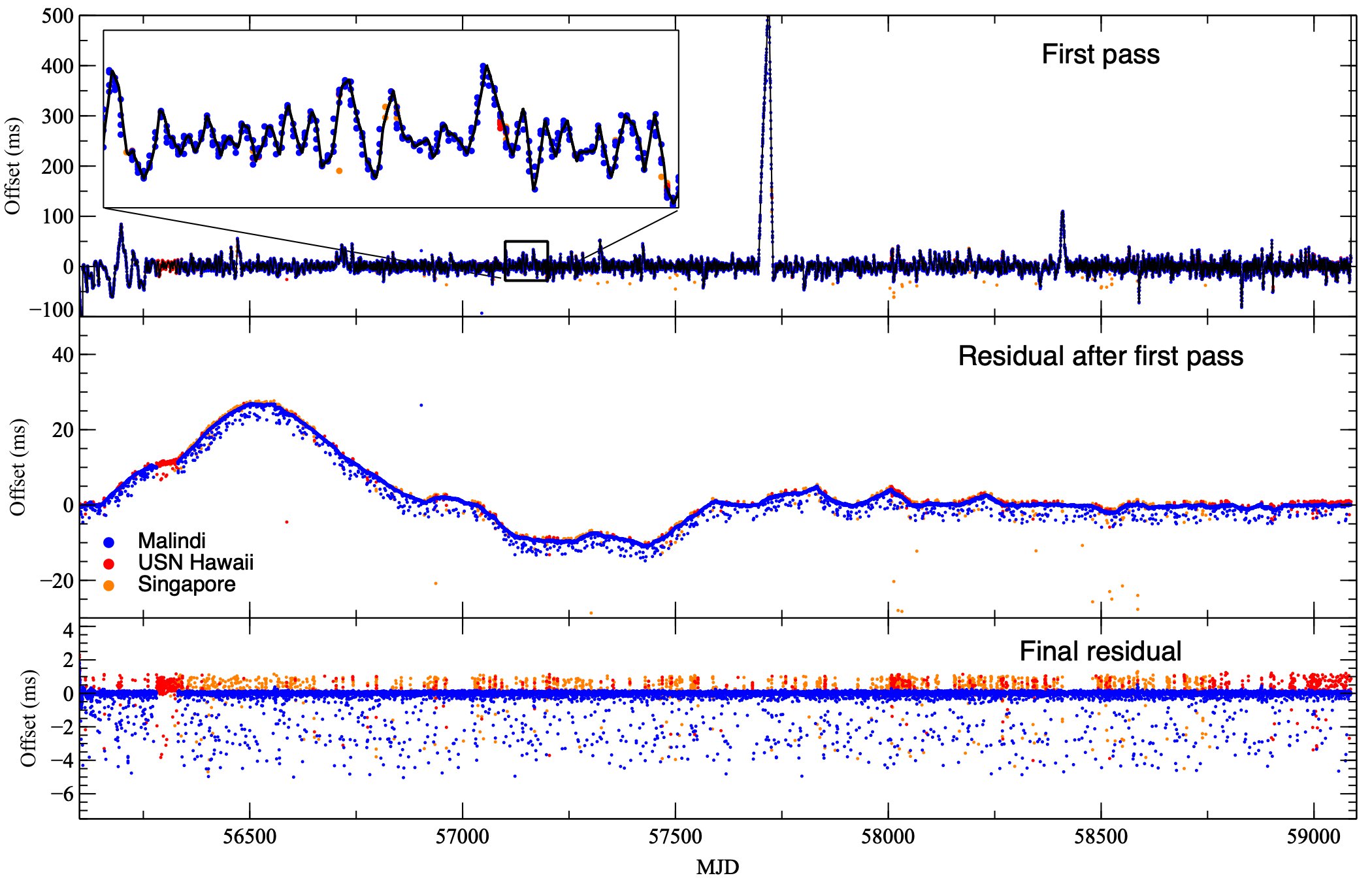}
\caption{Time history of the drift of the spacecraft clock.
Offsets are color-coded to indicate different ground stations, as shown in the legend.
(Top) Spacecraft clock offsets measured during ground passes, compared to the offset calculated with \eref{eq:correction} (in black; the inset shows MJDs 57100--57200);
(Middle) Residuals after correcting for the temperature model and adjusting for major trends. 
Note the large number of outlier measurements below a smooth long-term trend;
(Bottom) Residuals after eliminating the long-term trend with a robust spline fit
\label{fig:offset}}
\end{figure*}
The observable we want to measure here is the real TCXO frequency $f_0$, which is the reference for time tagging.

The \textit{clock divisor} $D(t)$ is a commanded value, and can be changed as needed by the ground station.
The spacecraft 1-PPS tick count rolls over when $D$ clock cycles have occurred.
The ``spacecraft time'' can be thus obtained:
\begin{equation}\label{eq:spacetime}
   \mathcal{T}(t) = \int_0^t {\frac{f_0(t^\prime)}{D(t^\prime)} d t^{\prime}}
\end{equation}

Periodic measurements from the ground station return the difference between the spacecraft time $\mathcal{T}_i$ and UT time $t_i$:
\begin{equation}\label{eq:offset}
    d(t_i) = \mathcal{T}_i - t_i + \epsilon
\end{equation}
where $\epsilon$ is an error term that we need to estimate. 
The mission operations center (MOC) removes known biases and delays, which we exclude from the formula above (but whose uncertainty we include in the error term).

Let us for a moment assume that we know the function $d(t)$ continuously and that we can ignore the error term. 
Using Eqs.~\ref{eq:spacetime} and \ref{eq:offset}, we get an estimate of the instantaneous spacecraft clock frequency
\begin{eqnarray}\label{eq:spacefreqcont}
  f_0(t) =& \dot{\mathcal{T}}(t) D(t) \nonumber\\
         =& [\dot{d}(t) + 1] D(t)
\end{eqnarray}
where the dotted variables indicate time derivatives. 
Considering that $\dot{d}(t)$ is measured in discrete intervals, the \eref{eq:spacefreqcont} can be approximated as
\begin{eqnarray}\label{eq:spacefreq}
  \bar{f}_0(t) \approx& \left[ \frac{\Delta{d}(t)}{\Delta t} + 1\right] D(t) \\
  	     \approx& \left[ \frac{d(t_2)-d(t_1)}{t_2-t_1} + 1\right] D(t)
\end{eqnarray}
where $\bar{f}_0(t)$ in this case is the \textit{average} frequency between subsequent clock offset measurements $d(t_2)$ and $d(t_1)$.
Therefore, by measuring the clock offsets and using the known commanded divisor, we can estimate the frequency change of the clock over a given interval.
The right hand side of \eref{eq:spacefreq} is only based on measured and commanded quantities.
The left hand side can be regressed against other quantities averaged over the same time intervals.

For convenience, from now on we will make the calculations not directly using $f_0$, but rather its deviation from the nominal TCXO frequency of 24 MHz
\begin{equation}
\mathcal{F} (t) = \frac{\bar{f}_0 (t) - 24\cdot 10^6}{24\cdot 10^6}
\end{equation}

Now let us suppose that the TCXO frequency is a function of temperature (an imperfect temperature compensation) and time (clock aging, which makes the quartz crystals less ``elastic''; see, e.g., \citealt{vig_aging_1991}).
We seek a relation of the form
\begin{equation}\label{eq:model}
\mathcal{F} (t, T) = \mathcal{A}(T) + \mathcal{B}(t) + \mathcal{C}
\end{equation}
where $\mathcal{A}$ is only a function of temperature, $\mathcal{B}$ an aging law, unrelated to temperature, and $\mathcal{C}$ is a constant.

We model $\mathcal{B}$ with a function of the kind:
\begin{equation}\label{eq:modelB}
\mathcal{B}(t) = b_0 \ln{b_1 (t - t_0 + 1)} + b_2 \ln{b_3 (t - t_0 + 1)}
\end{equation}
This particular form is motivated by the typical aging behaviors of crystal oscillators, with logarithmic changes of the frequency due to \textit{absorption} of contaminants and \textit{desorption} of crystal particles \citep{landsberg_logarithmic_1955,vig_aging_1991}.
In reality, we do not know if these exact processes are driving the aging curve in \fref{fig:f0ppm}, or if there are other processes involved. But this is the simplest physically-motivated model that is able to reproduce the full aging curve, as we will see later.
We favor this approach with respect to other approaches that are able to fit the data but need many more logarithmic components and are based on even weaker physical motivation \citep[e.g.][]{su_novel_1996}.

For $\mathcal{A}$, we use a quadratic function of the temperature:
\begin{equation}\label{eq:modelA}
\mathcal{A}(T) = a_0 (T - T_0) + a_1 (T - T_0)^2
\end{equation}
where $a_i$ are the fit coefficients.

In practice, we split the constant $\mathcal{C}$ into two parameters $c$ and $e$ that we used to fit separately for $\mathcal{A}$ and $\mathcal{B}$.

\begin{figure}
\includegraphics[width=\linewidth]{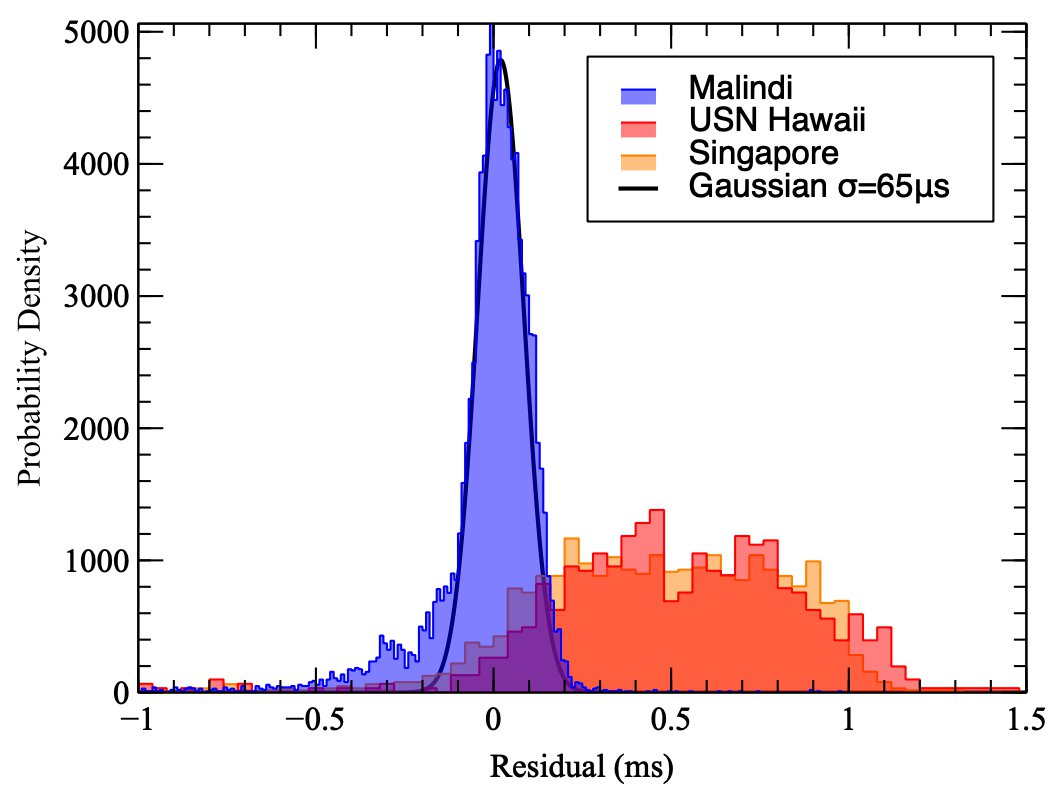}
\caption{Histogram showing the scatter of detrended residuals for the three ground stations. Note the much better performance of Malindi with respect to the other ground stations, with a scatter of  $\sigma\approx$65\us and an additional ``tail'' of outlier measurements.
The ``flat'' distribution of Singapore and Hawaii offsets is due to the way these ground stations provide clock offset measurements, truncated to 1 ms precision.
Since values are truncated and not rounded, the offset is systematically shifted in one direction.\label{fig:offset_detrend}}
\end{figure}
%After determining the best-fit values for the parameters $a_i$ and $b_i$ using averaged quantities, we go back to the beginning, and calculate numerically the expected clock offset using \eref{eq:offset}.

To calculate  the expected clock offsets (\eref{eq:offset}) we iterate the best-fit values for parameters $a_i$ and $b_i$, above, eliminating the effects of the aging law from the temperature law and vice versa.

The temperature of the TCXO is measured every $\sim10$\,s and recorded as part of the mission housekeeping information, so there are typically thousands of temperature measurements between ground segment passes.
Given a set of temperature measurements $T_k$ for times $t_k$, the numerical integral becomes
\begin{equation}\label{eq:correction}
\Delta d(t_i) = \sum^{i}_{k=0}  \left\lbrace\left[1 + \mathcal{F}(t_k, T_k)\right] \frac{24\cdot 10^6}{D_k} - 1\right\rbrace (t_{k + 1} - t_{k})
\end{equation}
where $D_k$ is the value of the divisor at time $t_k$. We can then model these offsets with a continuous function (e.g., a spline fit or a more robust interpolation) and compare them with the measured offsets.

The comparison between the measured offsets and the ones obtained in \eref{eq:correction} shows two very distinct problems with the clock offset measurements from the ground stations (see Figures~\ref{fig:offset} and~\ref{fig:offset_detrend}): 
\begin{itemize}
\item Only the Malindi station has a symmetric scatter around the best solution, while the Singapore and USN Hawaii are systematically overestimating the offset. 
This is a known effect, due to the fact that the latter two stations truncate their reference clock to 1\,ms precision. 
\item A large number of measurements, regardless of the ground station, underestimate the offset (the effect is so large that these bad measurements are often easy to flag just by comparing them with nearby measurements).
This is most likely due to processing delays within the spacecraft computer during the measurement, probably due to high CPU utilization.   Mission operations staff makes a best effort to
perform clock calibration during periods of expected low CPU
utilization, but this is goal is not always achieved.
\end{itemize}
These outlier clock measurements of both kinds can be easily flagged. After these measurements have been removed, we can go back to the beginning and fit Eqs.~\ref{eq:modelA} and \ref{eq:modelB} to the cleaned data. \\

In summary, the fit procedure is iterative, as follows:
\begin{enumerate}
\item  Fit the offset data to \eref{eq:modelA} with an initial value of the offset constant $c$
\item Fit \eref{eq:modelB} plus an offset $e$ a first time to the data \textit{after subtracting the results of 1.};
\item Flag all points at more than 0.02 parts-per-million (ppm) from the solution as outliers;
\item Fit \textit{only the linear term} of \eref{eq:modelA} to the data after subtracting the aging solution
\item Fit \textit{only the quadratic term} of \eref{eq:modelA} to the data after subtracting the aging solution and the linear term. This allows to constrain precisely $T_0$ thanks to the symmetry of the quadratic term.
\item Fit the full model A to the data one last time after subtracting the aging solution and fixing $T_0$. 
\item Fit \eref{eq:modelB} one last time on the temperature-detrended data.
\item Calculate the clock offset history using \eref{eq:correction}; flag outlier clock offset measurements.
\item Go back to point 1 and use the cleaned clock offset history to get more meaningful estimates of the local oscillator frequency. This time, calculate confidence intervals for the free parameters in \eref{eq:modelA} and \ref{eq:modelB}
\end{enumerate}

\begin{figure}
\gridline{\fig{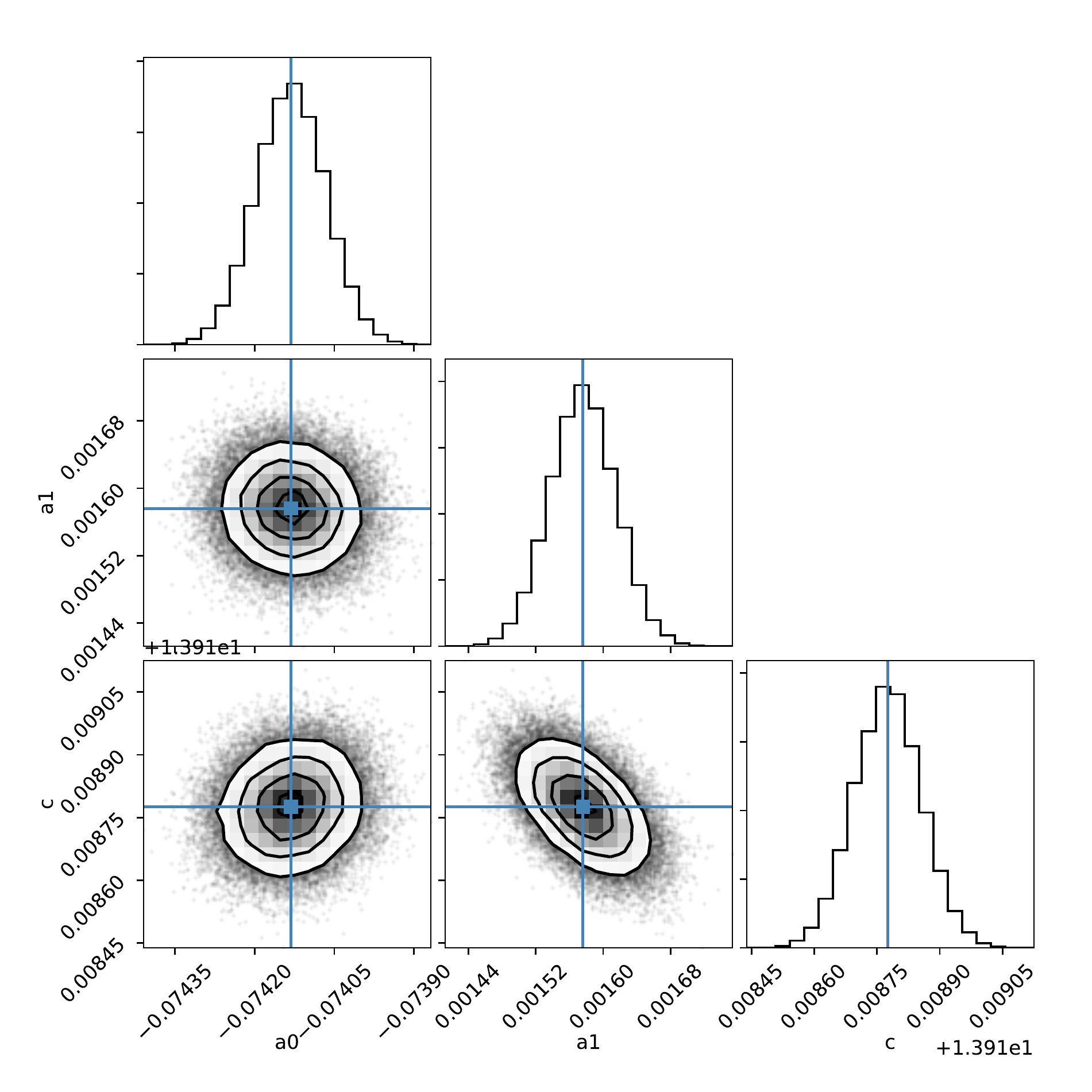}{\linewidth}{(a)}}
\caption{Modeling the imperfect temperature compensation of the spacecraft clock. Confidence intervals for the parameters of \eref{eq:modelA}, plotted using the \texttt{corner} library \citep{corner}.\label{fig:corner}}
\end{figure}

The results of the fit are presented in Table~\ref{tab:coeff} and Figures~\ref{fig:f0ppm}--\ref{fig:corner}.
We executed the fit with \texttt{lmfit} \citep{lmfit}. 
The confidence intervals on model A were calculated through a Monte Carlo Markov Chain using the \texttt{emcee} library \citep{emcee}, with uniform, unbounded priors, and the recommended burn-in and chain-length.
The high correlation of the parameters of model B (all but $e$) did not allow for meaningful confidence intervals with this method.

\begin{deluxetable}{ccCr}[b!]
\tablecaption{Best-fit coefficients in the clock correction model (Eqs.~\ref{eq:model}--\ref{eq:modelA}). 
\label{tab:coeff}}
\tablecolumns{6}
\tablenum{1}
\tablewidth{0pt}
\tablehead{
\colhead{Coefficient} &
\colhead{Unit} &
\colhead{Value}\\
}
\startdata
$T_0$ & C           & 13.440(25)\\
$t_0$ & MET         & 77509250\,\mathrm {(fixed) }\\
$c$ & ppm           & 13.91877(8) \\ 	 
$a_0$ & ppm / K     & -0.07413(7) \\
$a_1$ & ppm / K$^2$ & 0.00158(4)\\
$b_0$ & ppm         & 0.00829^*\\
$b_1$ & 1/Year      & 100.26^*\\
$b_2$ & ppm         & -0.2518^*\\
$b_3$ & 1/Year      & 0.0335^*\\
$e$   & ppm         & -0.0202(1)\\
\enddata
\tablecomments{$^*$ Uncertainties on the quantity $b_i$ are unreliable, due to the very high correlations between parameters. Therefore, we fixed them to the best fit in order to get a meaningful confidence interval on the offset parameter $e$, which is important to set the offset to the full model in \eref{eq:model}}
\end{deluxetable}

\subsection{Comparison with clock offsets}

\begin{figure}
\includegraphics[width=\linewidth]{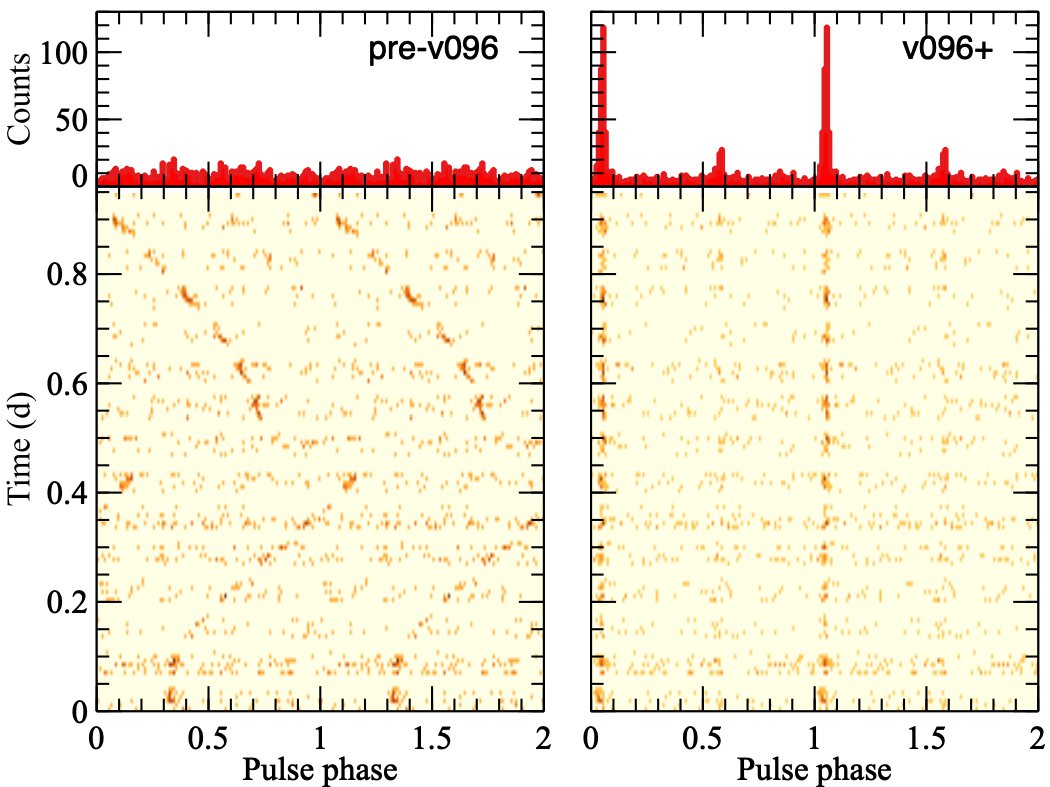}
\caption{Comparison of the timing performance obtained using the pre-v096 clock correction file (left) and that corrected with the technique described in this Paper (right). Shown is the time history of the pulse pulse profile of 
the rotation-powered millisecond pulsar B1937+21, folded on the ephemeris of \cite{arzoumanianNANOGrav11yearData2018} \label{fig:comparison}}.
\end{figure}

The results of the fitting procedure described above are shown in \fref{fig:offset}. 
The offsets are extremely well modeled throughout the full lifespan of \nustar. 
The difference between the calculated and observed offset is very small and produces an accumulated drift of $\sim200$\,ms in six years, which accounts for an error on $f_0$ of a part in a \textit{billion}. 
This long-term trend can be removed easily and is not shown in the figures.
Shorter-term trends remain, a slow drift of $\lesssim10$\,ms/yr.
The results are so precise that we can single out most of the points where the ground stations had a slow response, and we can easily take them out (see also \fref{fig:offset_detrend}). 
The remaining trends do not show any obvious correlation with temperature or other observables, but they can be interpolated with a spline.
We can now track the clock offsets on timescales of seconds instead of hours, and this allows a correction of the event arrival times to an unprecedented precision. 
The comparison of the performance of the method without using the clock correction file, with the legacy clock correction, and with the new method, are shown in \fref{fig:comparison}.

\section{Discussion: clock stability and precision}\label{sec:precision}

\subsection{Relative clock stability}
\begin{figure*}
\includegraphics[width=0.48\linewidth]{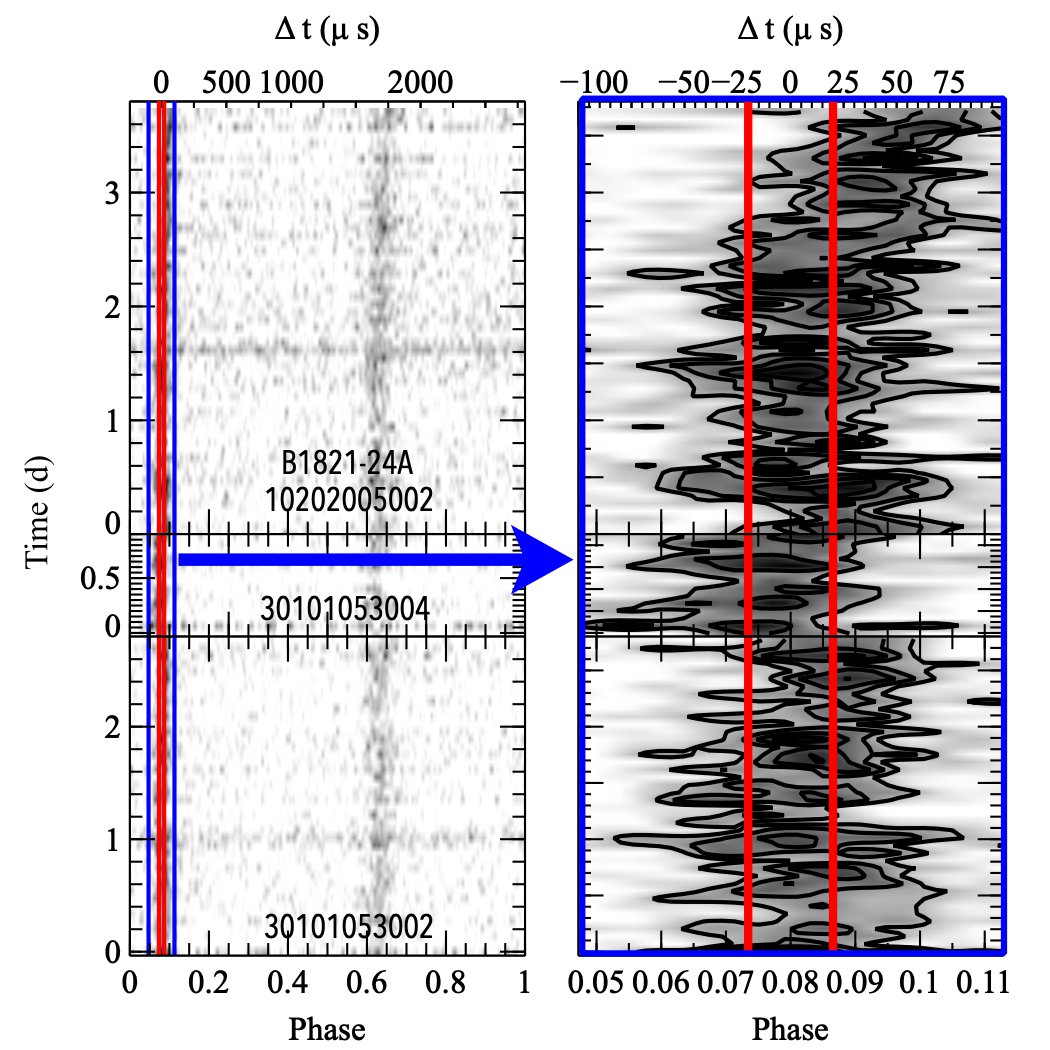}
\includegraphics[width=0.48\linewidth]{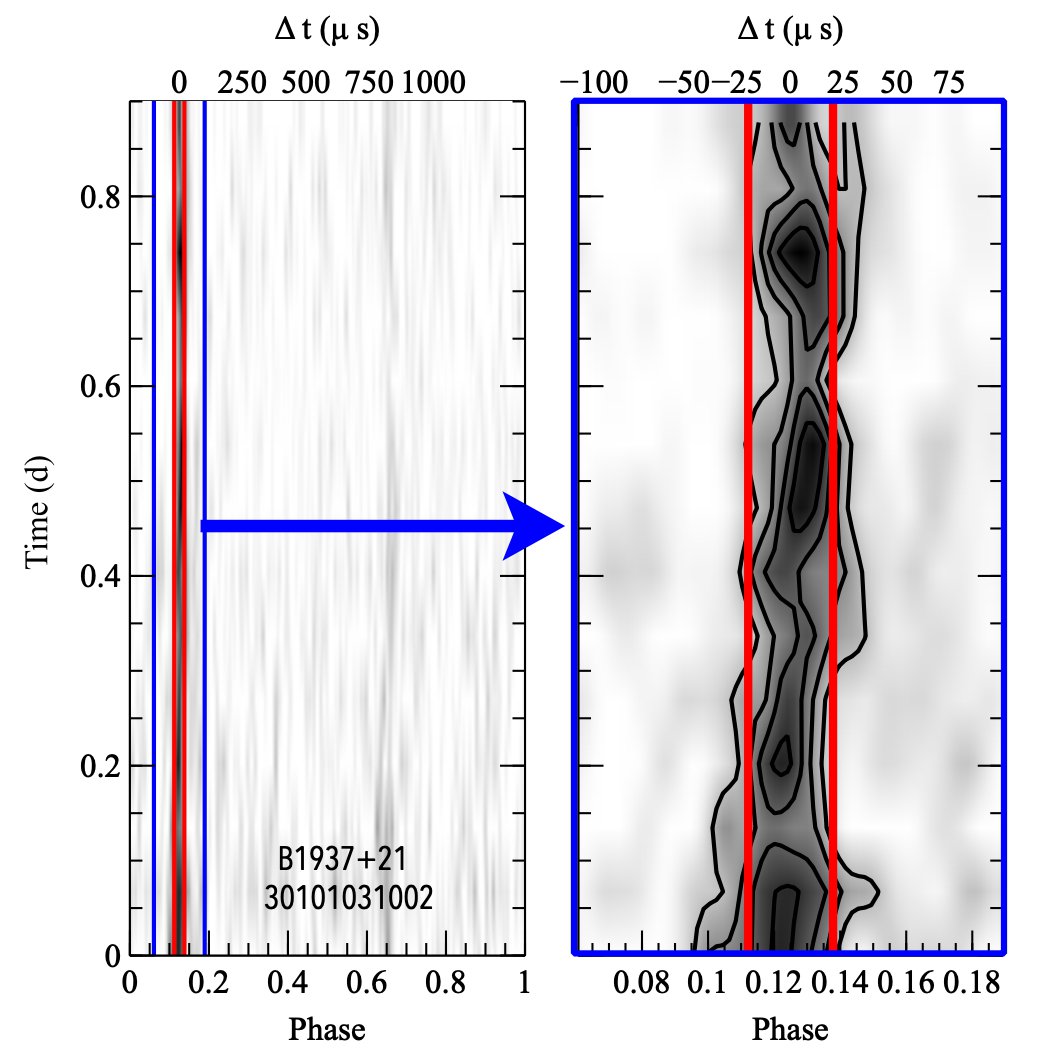}
\caption{Relative stability of the clock, shown through the phaseogram of three observations of PSR B1821-24A and one of PSR B1937+21.
To guide the eye, we plot in red a 40~$\us$ interval around the peak of the pulsar.
The peak of PSR B1821-24A is $\sim$20~$\us$ wide, and the additional wandering of the peak in timescales of $\sim1$ day is comparable to it.
The peak of PSR B1937+21 is thinner, and it stays inside the $\pm20 \ \us$ interval during a 1-day long observation.
In longer observation, the effect of the uncertainty on ground station offset measurements becomes visible, and the pulse phase shifts by a larger amount, compatible with the $\sim$60~\us measured in \fref{fig:offset_detrend}
\label{fig:stab}}
\end{figure*}
Figure~\ref{fig:stab} shows the clock stability during an observation.
When we compare the width of the pulse profiles with those tabulated in the literature, they are within 20 \us of the previously published values.

To get a more quantitative estimate of the residual clock drift after the temperature correction, we use the available data from the pulsars PSR\,B1821-24A and PSR\,B1937+21, chosen for their fast rotation and sharp X-ray pulse profiles \citep{saitoDetectionMagnetosphericXRay1997,takahashiPulsedXRayEmission2001}.
We split the dataset into 97~min long segments (approximately an orbit),
and fold the photons in each segment using the X-ray timing solution from the \nicer mission \citep{arzoumanian2014}, calculated by \citet{denevaLargeHighprecisionXRay2019}. 
We calculate the position of the peak in each data segment using the \texttt{fftfit} algorithm \citep{taylorPulsarTimingRelativistic1992}.
Finally, we measure the r.m.s. variation of the pulse phase around the median value\footnote{We expect the median value to be different from zero, due to the uncertainty in the ground station offset measurement.}.
We find the standard deviation of the pulse phase during each of the long observations of B1821-24A to be between 17.1 and 18.1~$\mu s$, compatible with the width of the main peak of the pulse profile as measured by \citet{denevaLargeHighprecisionXRay2019}.
Doing the same exercise with B1937+21, we find a much smaller jitter of 9~$\mu s$.
However, we caution that this observation is much shorter, and the number of points is very small.
This can produce deceptively small values of the standard deviation. 

% [[EXPAND]] The Allan standard deviation is represented in Fig.~\ref{fig:allan}

% \begin{figure*}
% \includegraphics[width=\textwidth]{allan_deviation.pdf}
% \caption{PRELIMINARY. 
% Allan standard deviation, calculated on long timescales from the residuals of the clock frequency model and on short timescales from timing the pulsar B1821-24A. 
% Both measurements give a best stability around 1 part per billion. \label{fig:allan}}
% \end{figure*}

\subsection{Absolute timing alignment}\label{sec:5ms}
As described in in Section~\ref{sec:precision}, the clock stability over $\sim$1 day of observations is $\sim$10~\us.
We now estimate the \textit{absolute} timing calibration, i.e. the precision of the \nustar clock with respect to a UTC reference.
This number is straightforward to measure by looking again at the clock offsets as measured from the ground station passages, and comparing them to the offsets calculated from the model in Section~\ref{sec:fvst}. 
Fig.~\ref{fig:comparison} shows how the additional residuals between the Malindi-measured offsets and the temperature model are easily modeled through a spline.
Once we do that, it becomes clear that the bulk of Malindi clock offsets are measured to a precision of $\sim$60~\us. 
Additionally, it becomes obvious that a large number of clock offset measurements are anomalous,
%off.
possibly the result of delays in the processing of the data from the spacecraft. 
%We attribute these bad measurements to delays in the processing of the data from the spacecraft. 
These outliers constitute a small fraction of the total measured offset and 
are easily separated from the ``good'' measurements.
%Being a relatively small number, these outliers can be easily separated from the ``good'' measurements.
What remains is a quasi-Gaussian distribution of clock offset measurements around zero, with standard deviation 65~$\us$, that defines the theoretical long-term reconstructed accuracy of the \nustar event timestamps.
%of \nustar.
% Question - is this the "theoretical accuracy" or the "[long-term] reconstructed accuracy" 

However, clock offsets measured by the ground station are not independent of electronic and instrumental delays, and might contain additional unmodeled biases.
The ultimate verification of the proper functioning of the \nustar clock correction procedures is to compare signals from fast pulsars measured with \nustar to those obtained by the \nicer mission that has a well-established absolute timing precision of $<$300~nanosec. 
We have selected a sample of fast ($<$70~ms) pulsars with sharp features in their pulse profiles for which there exist (quasi) simultaneous \nicer (0.2--12\,keV) observations.
This sample includes 3 rotation-powered pulsars, 1 recycled millisecond pulsar, and 4 accreting millisecond X-ray pulsars, and are described below. In total, we compared 15 simultaneous \nicer - \nustar observations obtained between April 2017 and February 2020.

\subsubsection{Source selection}
From the soft gamma-ray pulsar catalogue provided in Table 2 and Fig. 27 of  \citet{kuiper2015}, we selected a set of three rotation-powered pulsars satisfying our requirements: PSR B0531+21 (Crab pulsar; 33.5~ms), PSR J0205+6449 (65.7~ms) and PSR J2022+3842 (48.6~ms). The latter two pulsars are very weak at radio frequencies and are currently being timed by \nicer as part of an ongoing monitoring program \citep[see e.g. Sect. 4.1 of][for the method using time-of-arrival (TOA)  measurements]{kuiper2009}. Similarly, monthly radio timing ephemerides of the Crab pulsar are available from the Jodrell Bank Centre for Astrophysics \citep[][]{lyne1993}\footnote{see http://www.jb.man.ac.uk/~pulsar/crab.html}.

We also investigated the rotation powered (recycled) millisecond pulsars, PSR J0218+4232 (binary; P $\simeq$ 2.3 ms), PSR B1937+21 (isolated; P $\simeq$  1.55 ms) and PSR B1821-24A (isolated; P $\simeq$ 3.05 ms). These pulsars have hard non-thermal 
emission and (very) narrow pulses in their light curves \citep[see e.g.][]{kuiper2003,kuiper2004,gotthelf2017}, providing excellent timing calibration targets for the \nustar X-ray bands. 
However, concurrent \nicer and \nustar observations only exist for PSR B1821-24A. We used the \nicer observations of this pulsar performed during 2017 to generate an accurate timing model and  construct a high-statistics \nicer pulse profile to  compare with the deep ($\sim$155~ks) \nustar  observations performed in April and September 2017.

Finally, we searched for suitable sources amongst the Accreting millisecond X-ray Pulsars (AMXP) \citep[see e.g.][for an overview]{patruno2012}. %, that currently contain 19 members\footnote{Aq X-1, and the transitional AMXPs PSR B1023+0038 and XSS J12270}. 
Among the 19 AMXPs that went into outburst in the last few years, simultaneous \nicer and \nustar observations exist for the binary systems IGR J17379-3747 (P $\simeq$ 2.1 ms), IGR J17591-2342 (P $\simeq$ 1.9 ms), SAX J1808.4-3658 (P $\simeq$ 2.5 ms) and Swift J1756.9-2508 (P $\simeq$ 5.5 ms; 2 outbursts).
Timing models have been constructed for all selected AMXPs from the \nicer observations executed during the outburst of each source, which typically lasts a couple of weeks. 

\subsubsection{Data reduction}

\begin{figure}[t]
  \centering
  \includegraphics[width=\linewidth]{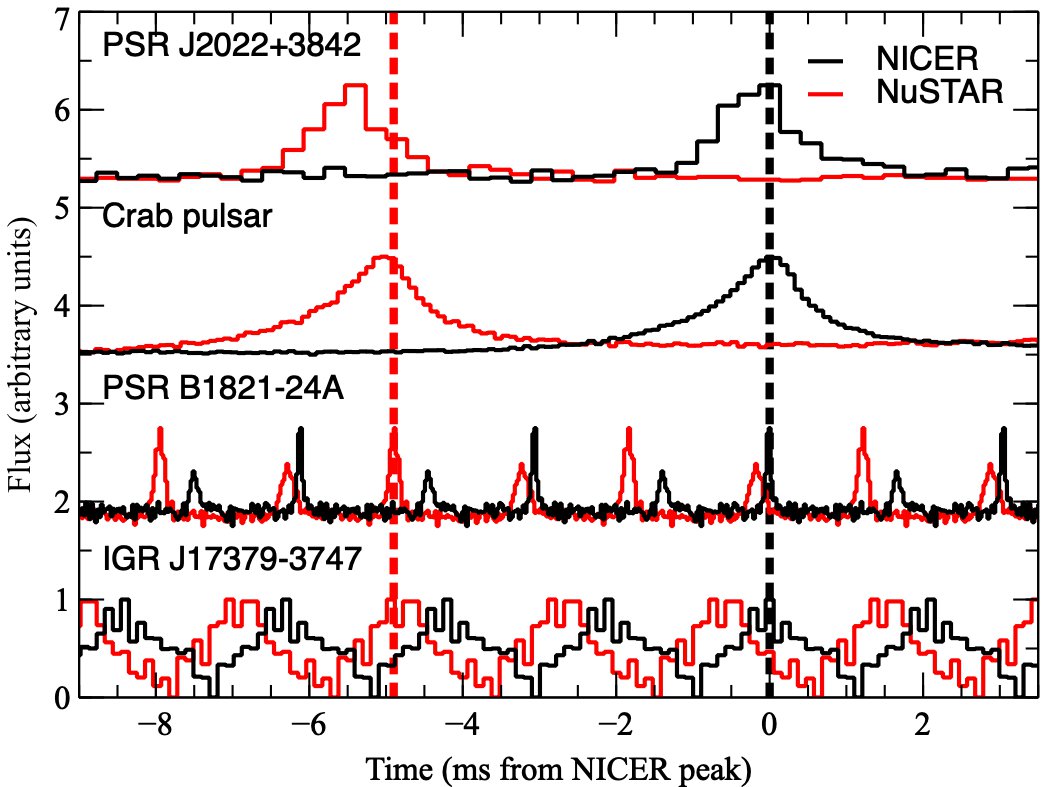}
  \caption{Comparison of arrival times of few selected pulsars quasi-simultaneously observed by \nicer and \nustar.
    \nustar data are systematically leading by $\sim5$\,ms with respect to \nicer
    Note that the fastest pulsars need more than one full rotation to complete these 5\,ms.
    Also note the slightly higher shift in PSR J2022+3842, that to a closer inspection turns out to be an outlier (see \fref{fig:abs_time_offsets})
  }
  \label{fig:fivemsshift}
\end{figure}
We reduced and analyzed the above pulsar sample according to standard procedures for \nustar and \nicer data. The photon arrival times at the satellite were corrected to the Solar System barycenter (in the TDB reference frame) using the same IDL code in order to avoid any potential differences between software tools. 
We used the most up-to-date location of the pulsars, the JPL DE200 (Crab) or DE405, and the spacecraft orbital ephemerides. For the selected AMXP sample we corrected the arrival times further for the orbital motion of the ms-pulsar in the binary system. 

The results were checked against the standard multi-mission barycenter FTOOL \texttt{barycorr} and found to agree at the few microseconds level.

To obtain comparison pulse profiles for each pulsar we selected photons from both instruments in the overlapping 3--10~keV bandpass.
Although the different responses of \nicer and \nustar bias the photons in this interval differently (\nicer towards softer photons and \nustar towards harder photons), we verified the effect on pulse shapes is negligible for our purposes.
We folded the corrected photon arrival times, $t$, into equal phase bins using same timing model for the both \nicer and \nustar data, according to the prescription,
\begin{equation} 
\phi(t)=\nu\cdot (t-t_0) + \frac{1}{2}\dot\nu\cdot (t-t_0)^2 + \frac{1}{6}\ddot\nu\cdot (t-t_0)^3. \label{eq:folding} 
\end{equation}
The timing model (ephemeris) in Eq. \ref{eq:folding} consists of the parameters ($\nu,\dot\nu,\ddot\nu,t_0$) representing the spin frequency, first order time derivative and second order time derivative of the frequency at epoch $t_0$, respectively. 

\subsubsection{Results}
\begin{figure}[t]
  \centering
  \includegraphics[width=\linewidth]{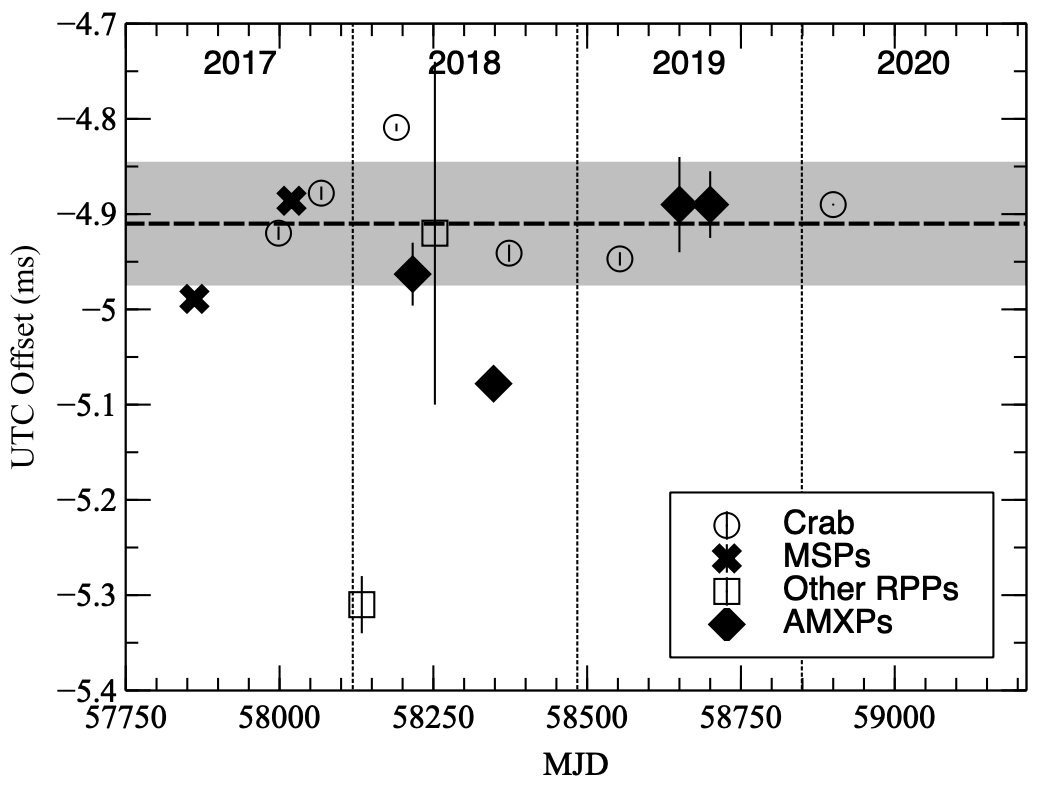}
  \caption{
    Distribution of time lags between \nustar and \nicer folded pulse profiles in the 3--10~keV energy band for a sample of 8 pulsar measured over April 2017 and February 2020. The pulse periods for these pulsars range from 1.9~ms to 65.7~ms. 
    Discarding the single outlier at MJD 58133, the weighted average time lag is -4.91~ms, as indicated by a dashed horizontal line. The one-sided $1\sigma$ confidence interval is $\sim68$~\us (grey area). 
%    enclosed by the dotted horizontal lines.
  }
  \label{fig:abs_time_offsets}
\end{figure}

As shown in \fref{fig:fivemsshift}, \nustar data are systematically leading \nicer data by $\sim5$\,ms.
To accurately measure this time lag and calibrate the absolute arrival time of the \nustar photons we cross-correlate the pulse profile for each pulsar against the reference \nicer profile to obtain their relative time lags (Fig. \ref{fig:abs_time_offsets}). The weighted average $\Delta T$ is $-4.92$\,ms with a $1\sigma$ uncertainty of $124\,\us$, or, after removing the PSR J2022+3842 outlier, $-4.91$\,ms with an uncertainty of $68\,\us$. The latter is notably compatible with the width of the distribution of the UTC values shown in Fig. \ref{fig:comparison}. Apparently, the \nustar clock runs about $4.91$ ms ahead of the baseline. 

\section{Clock correction file production}
\begin{figure*}[t]
  \centering
  \includegraphics[width=\linewidth]{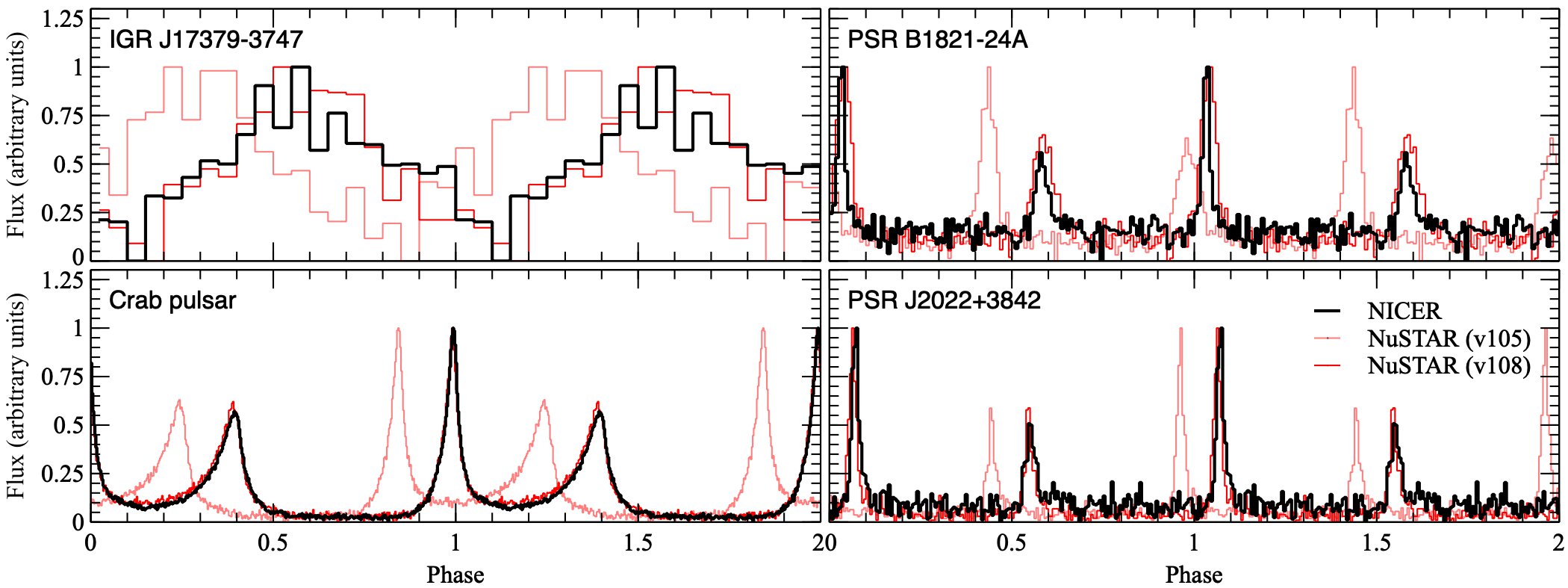}
  \caption{
    Application of the absolute timing correction to a sample of 4 different pulsars with pulse periods 2.1 ms (IGR J17379-3747), 3.05 ms (PSR B1821-24A), 33.5 ms (the Crab), and 48.6 ms (PSR J2022+3842) respectively. \nustar data are corrected using clock correction file v105 and v108 All profiles are derived for the 3--10 keV band.
  }
  \label{fig:corrections_pulsar_collage}
\end{figure*}
The standard \nustar clock files released by the SOC via the CALDB prior to version v096 were produced using an IDL interactive script 
to fit a spline through the raw clock offset measurements from the ground stations.
As mentioned previously, due to the rapid temperature-driven drifts of the spacecraft TCXO compared to the relatively sparse ground station passes, the clock correction was only reliable to $\sim$2\,ms \citep{madsen_calibration_2015}.

Since version v096, the distributed clock correction files are produced using the \texttt{nustar-clock-utils}\footnote{\url{https://github.com/nustar/nustar-clock-utils}}, based on the thermal model discussed in this paper. This procedure is automated and a new clock correction file with 
updated values is generated every $\sim2$ weeks.
Over the following versions, we improved the algorithm and the treatment of missing temperature measurements, bad ground station clock offset measurements, and so on.

The procedure to build the clock file works as follows:
(a) we update the latest clock offset measurements from all ground stations, the history of commanded divisor changes, and the temperature measurements;
(b) we take note of all time intervals where there are no temperature measurements for more than 10 minutes (e.g. because of reboots of the spacecraft software)
(c) for each continuous interval with no missing temperature measurements:
\begin{enumerate}
    \item the thermal model is applied the data (as in \fref{fig:offset}) and the residual clock offsets are calculated with respect to this thermal model (as in \fref{fig:offset});
    \item an initial model of the residual trends in clock offsets is created by making a robust polynomial interpolation with the outliers removed;
    \item the remaining residuals are smoothed using a median filter of 11 clock offset measurements;
    \item finally, the end points of the solution are adjusted to go to zero offset;
\end{enumerate}
(d) for each ``bad'' interval, having no temperature measurements, we use a straight interpolation of the solution between clock offsets and declare a fixed uncertainty of $\sim$ 1 ms;

The obtained correction has an overall median absolute deviation of $\sim 100\,{\rm \mu s}$.
Finally,
(e) we interpolate the correction using the offset and its gradient over a uniform grid of $\sim$3 points per \nustar orbit;
(f) we verify that the solution calculates the interpolation far from grid points with adequate accuracy;
(g) we apply a constant offset of 4.91\,ms, the best-fit cross-match from Section~\ref{sec:5ms};
(h) we create a clock correction file that contains four columns: the time at grid points, the clock offsets, the clock offset gradient, and the error bar calculated from the scatter of clock offsets in the days around the grid point.

Each new clock correction file is then tested as follows: we barycenter $\sim$45 test data sets obtained by the same number of \nustar observations of the Crab, PSR B1821-24A and PSR 1937+21. 
As periodic calibration observations of the Crab are executed, or any other good pulsar data are available, we add these to our test bench in order to track possible degradations of the clock performance.
We calculate TOAs using standard templates aligned with a standard X-ray observation (a previous verified \nicer or \nustar profile) and verify that they are consistently within $\pm100 \ \us$ of the expected arrival time throughout the mission time, with the exception of ``problematic'' intervals with no Malindi passes, missing temperature information, or other technical problems (see Appendix~\ref{sec:diagnostics}).
    
\section{Conclusions}
Over the years since the mission launch, we have conducted a deep study of \nustar's timing performance. 
We found the reference PPS signal produced by the spacecraft's TCXO oscillator, and used to time tag the events, is temperature-dependent. 
We characterized this temperature dependence and its change over the course of the mission, probably due to the aging of the TCXO.
In this Paper, we describe in detail this temperature dependence and how, correcting for it and using pulsar observations for a final alignment to UT time, we can achieve a sub-100\,$\us$ timing accuracy.
Finally, we describe how this temperature model is used to produce \nustar's clock correction files distributed with the \nustar Calibration database (CALDB).
We report on a $4.91$\,ms offset of the clock offset measured by the ground stations, now accounted for in the clock correction files. 

We conclude that, running the barycentering process with the official FTOOL \texttt{barycorr} using the clock correction files distributed by the \nustar SOC with the CALDB, \nustar event times can be trusted to the $\sim65\,\us$ level (1-$\sigma$) throughout the history of the mission, except for a few ``bad'' intervals whose list is continuously updated in the \nustar SOC page. 
The performance is continuously monitored using an ever-increasing set of test observations of multiple pulsars, in order to single out corner cases where the correction breaks down and/or promptly catch a possible degradation of the timing solution.

\acknowledgments
MB thanks the Fulbright Visiting Scholar Program for supporting a 9 month visit at Caltech.
DJW acknowledges financial support from STFC via an Ernest Rutherford fellowship.
The authors wish to thank: the \nustar X-ray Binaries WG, the Magnetars/RPP WG, and the Timing WG, for the hard work and fruitful discussions that led to, and improved, this work; Victoria Kaspi, Jill Burnham, Alessandro Riggio, Andrea Sanna, Paul Ray, Iulia Deneva, Franco Buffa and Alessio Trois for insightful discussions and suggestions.
They also wish to thank the referee for their comments, that helped improving the quality of this paper.
This research has made use of data and/or software provided by the High Energy Astrophysics Science Archive Research Center (HEASARC), which is a service of the Astrophysics Science Division at NASA/GSFC.

\vspace{5mm}
\facilities{NuSTAR,
            NICER,
            HEASARC}

\software{Nustar Clock Utils\footnote{\url{https://github.com/NuSTAR/nustar-clock-utils}},
          Stingray \citep{huppenkothen_stingray:_2016,huppenkothenStingrayModernPython2019},
          HENDRICS \citep{bachetti_hendrics:_2018},
          astropy \citep{price-whelan_astropy_2018},  
          PINT \citep{luo_pint:_2019},
          HEASoft/FTOOLS \citep{blackburnFTOOLSFITSData1995a,blackburnFTOOLSGeneralPackage1999},
          ATNF pulsar catalogue \citep{manchesterAustraliaTelescopeNational2005a},
          PRESTO \citep{ransomPRESTOPulsaRExploration2011},
          corner \citep{corner},
          emcee \citep{emcee},
          IDL\footnote{\url{https://www.harrisgeospatial.com/Software-Technology/IDL}}.
          Veusz\footnote{\url{https://veusz.github.io}},
          Holoviews\footnote{\url{http://holoviews.org}}
          }
\newpage
\appendix

\section{Example diagnostic plots for clock file testing}\label{sec:diagnostics}
To complement the creation of new clock files, we set up a test bed with an ever-increasing number (currently $\sim45$) of sample data sets containing at the moment:
\begin{itemize}
    \item four observations of PSR\,B1821-24A
    \item one observation each for PSR\,B1937+21, PSR\,J2022+3842, IGR\,J17379-3747
    \item 40 observations of the Crab pulsar
\end{itemize}

%The observations of the Crab only contain $\sim100,000$ photons for each obsID, as using the millions of photons in each would be an overkill and would considerably slow down the processing. We also did \textit{not} correct the Crab profiles for dead time, but created a dead-time affected reference profile and measured its precise alignment to a reference \nicer profile.
%This is also done to speed up the processing and allows the diagnostics to be ready in a few minutes instead of $\sim$tens.
%See Figures~\ref{fig:corrections_pulsar_collage}--\ref{fig:diagnostics} for the kind of diagnostic plots used in the process.
%See Figures~\ref{fig:corrections_pulsar_collage}--\ref{fig:diagnostics} for the kind of diagnostic plots used in the process.
%
For the Crab observations, we restricted the number of photons to a manageable level (100,000) for each observation, more then sufficient to precisely measure the alignment with the reference profile. 
However, dead time correction would require using all the photons in the observation.
Therefore, we created a reference dead-time-affected profile by cross-correlating its dead-time-\textit{corrected} analogous to the \nicer profile once and for all.
This allowed to have a robust dead-time-affected reference profile to be used with fast checks using only 100,000 photons per observation.

\begin{figure*}[t]
  \centering
  \includegraphics[width=0.7\linewidth]{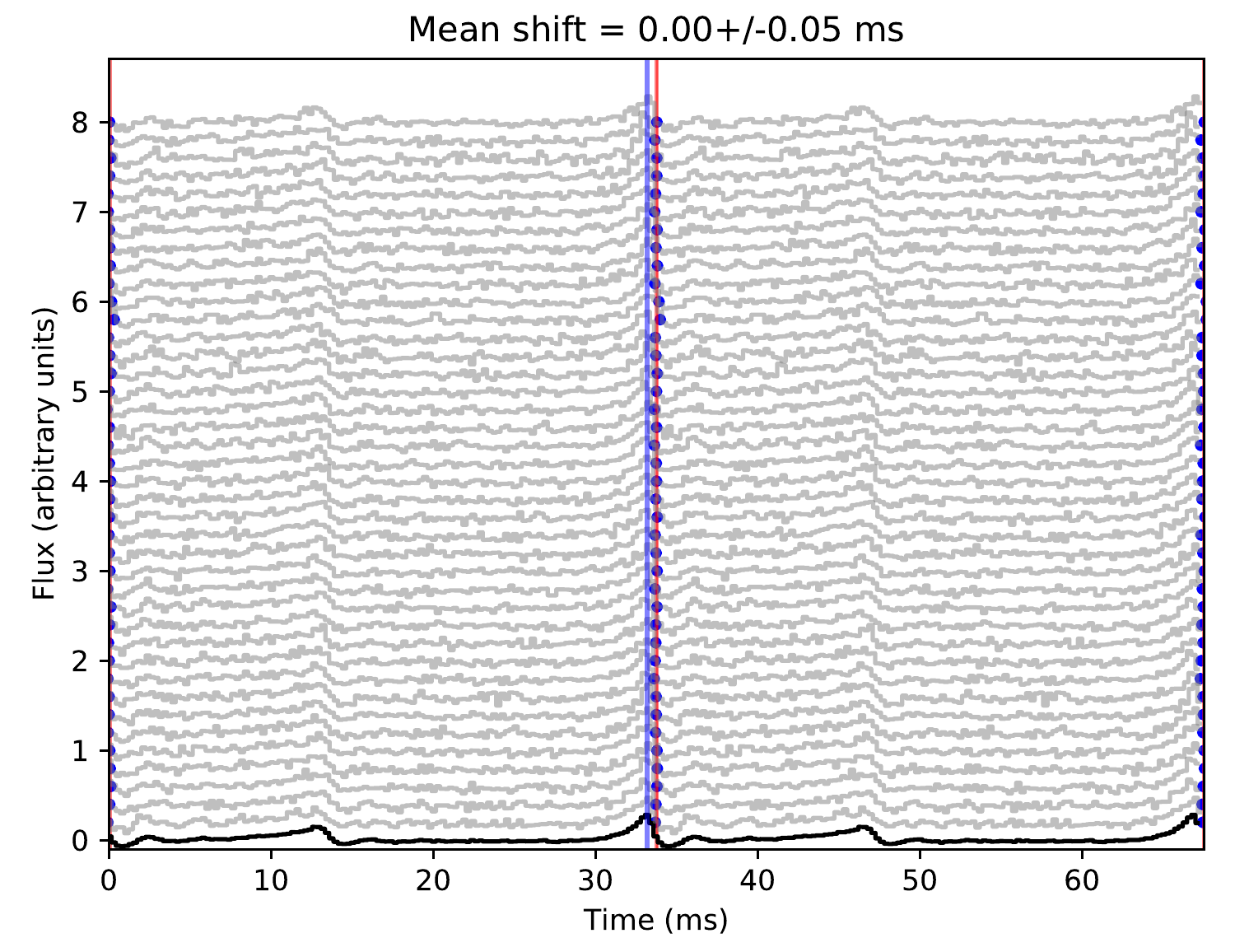}
  \caption{
      Diagnostic plot from the \texttt{nustar-clock-utils} showing the alignment of the Crab pulse over the history of \nustar observations. 
      The pulses are not dead-time corrected (hence the distortion of the profile) for performance reasons, but the template pulse was aligned so that its dead-time corrected version coincided with the \nicer profile $\pm100\us$.
  }
  \label{fig:diagnostics}
\end{figure*}

Figures~\ref{fig:corrections_pulsar_collage}--\ref{fig:diagnostics} show examples of the diagnostic plots used in the process.

\section{Pulsar searches with \nustar: known issues}
\nustar's detection chain has a high nominal time resolution and, as we demonstrated in this paper, allows an absolute timing precision of better than $\sim100\us$ in most observations. 
This allows, in principle, to search for very fast pulsars in \nustar data, with frequencies well above the physical limits for NS rotation.
However, there are a few known issues that people using \nustar for pulsar searches should keep in mind.
In this Appendix we will rapidly go through them.

The first, obvious, issue is dead time.
For the purpose of this Section, it is sufficient to note that \nustar's 2.5-ms dead time produces a frequency-dependent distortion of the PDF response.
This means that pulsar searches will have to use a frequency-dependent detection threshold or try to correct the shape of the PDS.
The problem is treated extensively in other papers \citep[e.g.][]{vanderklisFourierTechniquesXray1989,zhangDeadTimeModificationsFast1995}.
The presence of two identical detectors in \nustar allows to partly work around the problem \citep{bachetti_no_2015,bachettiNoTimeDead2018}.

All \nustar observations are executed in \textit{charge pump} mode (CPMODE; \citealt{miyasakaDevelopmentCadmiumTelluride2009}).
\nustar's detectors accumulate charges continuously due to internal (leakage) currents and other sources of electronic noise as well as the charge deposited by incident X-rays. To prevent the saturation of the readout electronics, a clock-synchronized feedback circuit removes this additional charge every 1.123\,ms (spacecraft time) in a ``CPMODE reset". The instrument experiences 30-40 \us of deadtime every time this happens.

A high-frequency pulsation search can often ``detect'' these resets as a strong oscillation close to $\sim 1/1.123s = 890.5$\,Hz and/or possible harmonics and aliases of this frequency. The actual frequency can change slightly based on temperature; also, being in spacecraft time, it sometimes goes undetected after barycentering. It is easy to single out the harmonics, as they are exact multiples of the fundamental. The aliases are just slightly more tricky, as they can represent non-obvious reflections of any harmonics about the Nyquist frequency.  They can be ruled out by changing the Nyquist frequency itself by using a light curve sampled at a different rate: if the feature does not change frequency, one can safely rule out that the feature is an alias of the 890\,Hz frequency.

In very early observations (prior to August 2012), there was an additional source of periodic dead time from housekeeping operations being run every 1, 4, and 8 seconds. A PDS of these early observations will typically show strong features at 0.125, 0.25, and all integer frequencies up to $\sim30$\,Hz.

\bibliography{nustartiming}
\bibstyle{aasjournal}

\end{document}